\documentclass[journal]{IEEEtran}

\usepackage{tikz}
\usepackage{textcomp}
\usepackage{hyperref}

\newcommand\copyrighttext{%
  \footnotesize \textcopyright {2018 IEEE. Personal use of this material is permitted. Permission from IEEE must be obtained for all other uses, in any current or future
  media, including reprinting/republishing this material for advertising or promotional purposes, creating new collective works, for resale or redistribution to servers or
  lists, or reuse of any copyrighted component of this work in other works. DOI: \href{<https://doi.org/10.1109/TPDS.2018.2814567>}{10.1109/TPDS.2018.2814567}} 
   }
    
\newcommand\copyrightnotice{%
\begin{tikzpicture}[remember picture,overlay]
\node[anchor=south,yshift=10pt] at (current page.south) {\fbox{\parbox{\dimexpr\textwidth-\fboxsep-\fboxrule\relax}{\copyrighttext}}};
\end{tikzpicture}%
}

\usepackage[linesnumbered,ruled]{algorithm2e}
\usepackage{graphics}
\usepackage{setspace}
\usepackage{subfigure}
\usepackage{color,soul}

\soulregister\cite7
\soulregister\ref7
\soulregister\pageref7 
\usepackage{amssymb,array,url}

\usepackage[T1]{fontenc}
\usepackage{listings}
 \SetKwFor{Case}{case}{}{}%

  \SetKwFor{Case}{case}{}{}%
  \usepackage{tabularx,ragged2e,booktabs,caption,setspace}
\usepackage[numbered]{matlab-prettifier}

\lstdefinestyle{mymatstyle}{%
  style=Matlab-editor,
  basicstyle=\mlttfamily,
  frame=leftline,
  numberstyle=\scriptsize,
  xleftmargin=1.8em,
}

\newcolumntype{L}[1]{>{\raggedright\let\newline\\\arraybackslash\hspace{0pt}}m{#1}}
\newcolumntype{C}[1]{>{\centering\let\newline\\\arraybackslash\hspace{0pt}}m{#1}}
\newcolumntype{R}[1]{>{\raggedleft\let\newline\\\arraybackslash\hspace{0pt}}m{#1}}

\usepackage{listings}
\newcommand{\quotes}[1]{``#1''}
 
\usepackage{longtable}
 
\usepackage{cite}
\usepackage{amsmath}
\usepackage{algorithmic}

\usepackage{array}

\ifCLASSOPTIONcompsoc
  \usepackage[caption=false,font=normalsize,labelfont=sf,textfont=sf]{subfig}
  \usepackage[caption=false,font=footnotesize]{subfig}
\fi
\usepackage{fixltx2e}

\usepackage{stfloats}
\usepackage{url}

\newlength\mylength
\begin{document}
\makeatletter
\let\old@ps@headings\ps@headings
\let\old@ps@IEEEtitlepagestyle\ps@IEEEtitlepagestyle
\def\confheader#1{%
  \def\ps@headings{%
    \old@ps@headings%
    \def\@oddhead{\strut\hfill#1\hfill\strut}%
    \def\@evenhead{\strut\hfill#1\hfill\strut}%
  }%
  \def\ps@IEEEtitlepagestyle{%
    \old@ps@IEEEtitlepagestyle%
    \def\@oddhead{\strut\hfill#1\hfill\strut}%
    \def\@evenhead{\strut\hfill#1\hfill\strut}%
  }%
  \ps@headings%
}
\makeatother

\confheader{%
 \begin{minipage}{\textwidth}
 \centering
 \tiny{ This article has been accepted for publication in a future issue of this journal, but has not been fully edited. 
    Content may change prior to final publication. Citation information: DOI 10.1109/TPDS.2018.2814567, IEEE Transactions on Parallel and Distributed Systems. 
    ->1045-9219 (c) 2018 IEEE. Translations and content mining are permitted for academic research only. Personal use is also permitted, but republication/redistribution requires IEEE permission. See <http://www.ieee.org/publications\_standards/publications/rights/index.html> for more information.}
    \end{minipage}
}

%
\title{TripleID-Q:  RDF Query  Processing  Framework using  GPU}
%
%
%

\author{Chantana~Chantrapornchai 
      Chidchanok Choksuchat  
\thanks{Chantana Chantrapornchai is with the Department
of  Computer Engineering, Faculty of Engineering, Kasetsart University, Bangkok, Thailand
e-mail: fengcnc@ku.ac.th}
\thanks{ Chidchanok Choksuchat  affiliated  with  Silpakorn University, Nakhon Pathom, Thailand, and
currently  she works  for Information and Communication Technology,
Faculty of Science, Prince of Songkla University, 
Songkhla, Thailand, email: chidchanok.ch@psu.ac.th}
}

\maketitle
\copyrightnotice

\begin{abstract}
 Resource Description Framework (RDF) data  represents  information linkage around the Internet. 
It uses Internationalized Resources Identifier (IRI)  which can be referred to 
external information.  Typically, an RDF data is serialized as a large text file which contains  millions of relationships.
 In this work, we propose a framework based on  \emph{TripleID-Q},  for  query processing of  large RDF data in a GPU.   The key elements of the framework are 1) a compact representation 
suitable for a
 Graphics Processing Unit (GPU) and 2) its simple representation conversion method  which optimizes the preprocessing overhead.  
Together with the framework, we   propose 
parallel algorithms  which utilize
 thousands of GPU threads to look for specific data for a given query as well as 
to perform basic query operations such as union, join, and filter.
The TripleID  representation is smaller than the original representation 3-4 times.
Querying from TripleID  using  a GPU  is    up to  108 times faster than using  the traditional RDF tool.
The speedup can be more than 1,000 times over the traditional  RDF store when processing a complex query with union  and join of many subqueries. 
 

\end{abstract}

\begin{IEEEkeywords}
Query processing, Parallel processing, Entailment,TripleID, GPU, RDF.
\end{IEEEkeywords}

%
\IEEEpeerreviewmaketitle

\section{Introduction}
Linked data  \cite{linkeddata} utilize   web resources  to connect related data around the Internet.  They contain common data such as DBpedia  \cite{dbpedia},
 biomedical data \cite{bioportal}, geographical features data \cite{geoname},  etc.
These linked  data are represented in Resource Description Framework (RDF) \cite{RDF}
which is a standard and common framework  to share and reuse data across the Internet.
RDF  data 
contain   relationships, each of which is in a triple statement:  subject, predicate, and object. \emph{subject} denotes the resource, \emph{predicate} shows the property of the subject and \emph{object} is the value of the property.
 Each of these, subject, predicate, and object, is usually an Internationalized   Resource Identifier (IRI), which is   a very long string.    RDF  data contain millions of triple statements which result  in a significant data size. Thus,  it is time consuming to load and queries such those million triples.

 With  current parallel technology and architecture, it is possible to utilize multi-threading  to perform such tasks to speedup the overall processing time.
Current  architecture  has been advanced allowing it  to process applications using multi-threading on many cores. Multi-threading can be in the form of high-level concurrency using Java Executor Service \cite{java8, Corco} or  low-level CPU threads such as OpenMP or pthreads for a multi-core or many core computer.
GPUs are one of such hardware platforms that contain   many thousand cores.
Due to  its inexpensive cost, it becomes a cost-effective platform to gain  high-speed processing, especially for imaging and graphic applications. 
Nowadays, a  GPU has  been used for general-purpose computing in many other application areas \cite{CUDA}. 
However, to use  the GPU, applications must be designed properly to support the GPU architecture.

Though there are many open source tools for  querying RDF data  such as Redland \cite{Redland}, RDFlib\cite{RDFlib}, RDFsh \cite{RDFsh} , HDT \cite{FMPGPA13} etc.,  
which are  easy to use, 
some of them are implemented in scripting languages which usually consume lots of time to load data, to create  the internal  representation as well as  to query the model when the data become very large.  Some are  the libraries interfacing with C or Java with a complex data structure, making it difficult to port to utilize GPU  to speedup the processing.
Free community version can process limited number of triples (around 20 millions)   \cite{stardog}.
The well-known open source one such  Virtuoso  \cite{virtuoso} can support larger number of triples but do not support the use of GPUs. Blazegraph \cite{blazegraph} is a high-performance graph database supporting Semantic Web (RDF) and SPARQL query on CPUs and GPUs with Java language but it also is not offered as an open-source or community-edition products on GPUs' version.

To utilize  a GPU  for  query processing,  we have to consider two  main aspects:
 the GPU architecture and  the nature of the RDF query processing.
For the first issue, a contemporary GPU have thousand  cores supporting many concurrent threads.   All these  threads
share the GPU  memories.
The GPU memory size is limited and the data must be transferred to the GPU memory before these threads can start computing.

To process  an RDF query, all RDF data must be entirely loaded 
 and stored in certain data structure. 
The aforementioned RDF libraries use graphs and  heap storages to store RDF data. 
Some framework creates  indexes  for fast processing such as Header Dictionary Triple (HDT) which extracts common terms and creates dictionary as well as index triples by subject \cite{FMPGPA13}. This  format compresses the original RDF data very well. 
However, the implementation of these above data structure  mostly  are based on a list iterator, or recursive pointer. They contains deep pointers  which are complex to load data  GPU memory and let the threads to work on.
 
To process queries using GPU threads, data must be transformed 
into  a proper form. The format  should be compact so that all million triples can reside in a GPU memory.
Also, the data structure should allow  threads to look for proper relations with a high degree of parallelism.

Our  research goal is  to speedup  large RDF query processing using a GPU.  
In order to achieve this goal, the following  subproblems  are investigated.
\begin{itemize}
\item How to design the compact representation for RDF data that is suitable for the GPU memory layout.
\item  Decide the information that needs to be inside the GPU memory for processing.  
\item  How to utilize the GPU threads   for concurrent processing.
\item   How to integrate the tasks performed by a GPU and CPU to obtain   final query results.
\end{itemize}

 To address the above issues,
we propose a simplified format, \emph{TripleID},  which is a transformed representation to
encode the RDF data into unique IDs. The  conversion to TripleID  can be done in linear time.
Such a file is small and  can be easily loaded to the  GPU memory. The data are kept in GPU memory as long as needed.

 We adapt the search algorithm to utilize GPU threads to look for specific data according to a user query.
The found data are returned to the GPU host and then mapped back to the corresponding name. The CPU side manages how to store, and select the returned data properly. It sends  new data to the GPU   for the next   search. A CPU and GPU    interact with each other depending on query operations such as union, intersection, or join.
 To lookup  TripleID, the GPU threads are invoked. There is no need to transfer  data to the GPU memory again.    Some data returned from the GPU  may be removed due to redundancy and may be merged with previous returned results. CUDA Merge-Join and Thrust libraries are used to speedup the processing of intermediate results \cite{CUDA,thrust}.

 Such  framework,  \emph{TripleID-Q},  can be used for querying  RDF data.
In the experiments, we demonstrate the use of the framework starting  from taking  RDF data in the triple form (N-Triples and/or N3) \cite{ntriple}
and converts them into TripleID.
Then, all  IDs are loaded to the GPU memory. 
 The converted TripleID files are 2-4 times smaller than the original NT files and
 the  conversion time to TripleID is  3 times  faster  than other well-known representation.
The TripleID loading time is faster than   the original  NT file loading time and common RDF store loading time.
The framework can process  a  simple query faster than traditional RDF library.
  Especially for the complex queries with lots of intermediate results,   1,000-time speed up or more can be obtained compared to querying using  the traditional RDF store.

The outline of this paper is as follows: In Section \ref{sec:back},  background of RDF and  a GPUs as well as related work are presented.  
After that, our approach is presented in Section  \ref{sec:method}  which  includes data  representation for GPU search. 
Section \ref{sec:method2} gives an example of adopting the representation for different query operations.
In  Section \ref{sec:exp}, the experiments  comparing the data size reduction and conversion time are presented. 
 The query   processing time of our approach is  compared with that of  the traditional tools.  
Section \ref{sec:conc} concludes the paper and describes the future work.

\section{Backgrounds}
\label{sec:back}
 
 We  introduce the Resource Description Framework (RDF), a Graphic Processing Unit (GPU), then   related works.

 \subsection{Resource Description Framework (RDF)}
Resource Description Framework (RDF) is a common format used to describe  data in a  relation form.
It is represented in  a triple form, (\emph{subject, predicate,object}) where  each term is usually a Internationalized   Resource Identifier (IRI) which  can be linked to another web resource \cite{W3CRDF}. 

An example of the RDF triple is shown as:

{\small\singlespace
\begin{verbatim}
<http://www.owl-ontologies.com/
     BiodiversityOntologyFull.owl#Air> 
<http://www.w3.org/2000/01/rdf-schema#subClassOf> 
<http://www.owl-ontologies.com/
    BiodiversityOntologyFull.owl#AbioticEntity>
\end{verbatim}}

The above triple implies  \texttt{Air} is a subclass of \texttt{AbioticEntity} based on RDFS vocabulary. 
{\small \url{<http://www.owl-ontologies.com/BiodiversityOntologyFull.owl#Air>}} is a subject,
{\small   \url{<http://www.w3.org/2000/01/rdf-schema#subClassOf>}} is a predicate, and
{\small   \url{<http://www.owl-ontologies.com/BiodiversityOntologyFull.owl#AbioticEntity>}}    
is an object. They all  are IRIs and
are obtained from  \emph{biomedical ontology} \cite{bioportal}.

Searching to the RDF data is  done   by the query language, SPARQL \cite{clark_sparql_2008}.  
 A SPARQL's SELECT statement is similar to  SQL SELECT statement.
A  given query can ask for subjects, predicates,  and/or objects of the triples. 
The query in Listing 1 contains two subqueries, asking to \quotes{find all authors of The Journal of Supercomputing}, adapted from \cite{vir2010}. \texttt{ ?authors} are variables whose values are the answers for the SELECT statement.  \texttt{dc} is an abbreviation prefix of \url{<http://purl.org/dc/elements/1.1/> } which  is a standard vocabulary resource from Dublin Core \cite{dc}.

 \begin{lstlisting}[style=mymatstyle, 
language=SPARQL,
showspaces=false,
basicstyle=\ttfamily,
commentstyle=\color{gray},
caption = {find all authors of The Journal of Supercomputing},
label = listing1]
PREFIX dc: <http://purl.org/dc/elements/1.1/> 
SELECT ?yr ?authors
WHERE {
?journal  dc:title 
   "The Journal of Supercomputing"^^ xsd:string .                                   
?journal dc:creator ?authors . }
\end{lstlisting}


If one would like to infer a subclass ( \texttt{rdfs:subClassOf}) between any two terms \texttt{?x,?z}. We can create two subqueries that
are connected via a temporary variable,  i.e.,
if $?x$ is a subclass of $?y$, and $?y$ is a subclass of $?z$, then $?x$ is a subclass of $?z$ \cite{_rdf_type} as shown in Listing 2.
 
  \begin{lstlisting}[style=mymatstyle, 
language=SPARQL,
showspaces=false,
basicstyle=\ttfamily,
commentstyle=\color{gray},
caption = {Subclass transitivity},
label = listing2]
SELECT ?x ?z
WHERE {
?x  rdfs:subClassOf ?y .    
?y  rdfs:subClassOf ?z .  } 
\end{lstlisting}


To process the above query, using  a traditional RDF tool, it is necessary  to load all triples into the memory. The triples are stored in data structures such as graph models. 
Each  subquery is  then processed and the results from each subquery are kept for merging.

\subsection{Graphics Processing Unit (GPU) and Compute Unified Device Architecture (CUDA)}

 A Graphics Processing Unit is  originally used to process graphics objects for display.  With the advanced  hardware,
they  contain  thousand cores which can be used to do any kind of general-purpose computations in parallel.
Though they have a lower clock speed than the   CPU, the thousand  cores can process faster if they are utilized  properly. 
 
In general,  a GPU, sometimes called \emph{ device},  resides in a computer, called \emph{host}.
 To utilize the GPU, a proper programming framework is   needed. Compute Unified Device Architecture (CUDA) is one of the commonly used framework supporting   an NVIDIA GPU   \cite{CUDA}. 
 In CUDA,  threads are organized as grids of thread blocks. Threads in a block are executed simultaneously. 

%

CUDA cores are grouped into Streaming Multiprocessor (SM). One GPU card contains 4-26 SMs.
    A GPU  has many types of memories  such as   \emph{local},  \emph{shared},  \emph{global} memories, etc.
 Global memory  can be accessed by all threads in all blocks while the shared memory   can be accessed by only threads in the same block.
  Global memory has the largest sizes, varying from 2G to 24GB depending on the card models.  Even though the access time is slower than that of shared memory, the shared memory usually has the size up to 112KB.
For general-purpose computing,  the global memory  is commonly utilized since it is the largest and and it can be both read and written. 
In some cases, for small frequently accessed data, the shared memory may be used.
The data from the global memory must be copied to shared memory  before accessing them.

Under this architecture, the GPU memory transfer latency can be an obstacle to improve the program execution time.
Algorithms that utilize the GPU must be designed in such a way that
the required data needs to be kept inside the GPU memory as long as possible to reduce the transfer time, thus reducing the whole execution time.

In our case, all RDF data must be 
transferred  to the GPU memory before the querying process can be done.    
Since RDF data is large, global memory is used to store all of them. Compacting them will  be  advantageous since more RDF data will be held.
 The search is performed by concurrent threads and the found triple positions are returned. Complex queries processing can also be done inside the GPU memory with proper data arrangement.
 
\subsection{Related work}

 Since we are interested in processing large RDF data using a GPU or a parallel platform.  Such a platform  has lots of computing nodes/cores  which can be advantageous for parallel processing. Also, the platform needs all data on the device's memory for processing while it has limited memory size.
 Thus, we study the previous works in two aspects: 1) utilization of a GPU  or  any parallel platform  for information processing 2) the advantage of compacting data for saving memory storage or splitting data for concurrent processing.
 
 \subsubsection{RDF processing with parallel platforms}
 


With the advancement of parallel platforms with many computing cores and  bigger memory, large information can be stored and processed inside the device. The information  processed can be in various forms such as database, large text files, or RDF data etc.
   He et al.  considered  speeding up relational database using the  GPU  \cite{He2008}. The authors  focused on designing data-parallel primitives such as   \emph{split}, \emph{merge}, \emph{map}, \emph{gather-scatter}, \emph{sort}, and \emph{join}, for  memory optimization. The main problem in GPU programming  is that the array in the GPU memory must be  allocated  before the GPU kernel is invoked.  They developed the \emph{lock-free} scheme for storing result outputs where  two phases are used: the first phase was to examine the total size of the results for the GPU memory allocation, the next step  was to perform the operation on the result array in the GPU. 
   Bre{\ss}  
et.al.\cite{Bre2014} proposed a workload optimization scheme, called probability outsourcing. They considered benchmarking of  4 database operations \emph{aggregate, select,sort, and join} across GPU devices.  The implementation is based on CUDA framework. 
 Groppe, et.al. focused on distributed merge join processing  for  RDF triples \cite{Groppe201}.  They used partitioned $B+$ tree for  indexed triples.  The indices were built using a cluster of 7  computers. Another concurrent technology available was  Java stream and multithreading where Corcoglioniti et.al.  \cite{Corco} proposed  a  library tool for process RDF data supporting   \emph{filter},\emph{aggregrate}, \emph{inference}, \emph{deduplication}. The tool processes the data in a pipeline fashion.

Some researchers were  interested in inferring knowledge from RDF data, called \emph{RDF Schema (RDFS) entailment}.
RDFS contains  a standard set of rules for an RDF vocabulary which  new relations can be inferred from. 
One of the motivated works to us was presented by Heino and Pan. The RDFS entailment was performed  on a cluster of CPUs with one device, (and subdevices) \cite{HJ2012}.  Their algorithm was implemented  using OpenCL while the RDF graph representation was used.  The steps of the entailment were similar to \cite{UJK2009} while there was a synchronization between  steps. The key concept was to remove duplicate items before sending the results back to the CPU to save the data transfer time and to compact the transfered data.
Liu et.al \cite{Liu2014} studied the problem of reasoning for RDF reasoning using streaming RDF triples over time. These reasoning rules can be implemented using several subqueries.
Makni  \cite{Mak2013}'s proposal  focused on social media data stream which can be  often changed. 

Table \ref{tab:compareprev} summarizes the previous work mentioned and compares them in the aspect of target tasks, representation and platform tested. The works in\cite{Bre2014, Groppe201, He2008} focus on relational database operations
while the work in \cite{Bre2014} targets at query plan optimization through various GPU devices.  The work in \cite{HJ2012, Corco} targets the RDF processing where entailment problem was considered in \cite{HJ2012} and the later work in \cite{Corco} presents Java library for RDF processing. Most of these work used hash table for speeding up the query while some utilizes indexing scheme such as $B+$ tree.
Our work in the last row, we consider the similar common operators with TripleID representation without spending time to generate indices. The compact representation allows GPU to process large number of triples as well as RDFS entailment.

\begin{table}\small
\setlength{\tabcolsep}{2pt}  
\begin{center}
\caption{Comparison of previous works in RDF processing schemes using parallel technology.}
\label{tab:compareprev}
 
\resizebox{\columnwidth}{!}{%
\begin{tabular}{|L{2cm} ||L{2cm} |C{1.5cm}|L{2cm}|}\hline
Previous works & Platforms & Representation & Target tasks \\\hline
He et al.  \cite{He2008}.  & GPU/CUDA&   N/A & Relational database operation: \emph{join, sort, gather-scatter, map} \\\hline
 Heino and Pan \cite{HJ2012} & A cluster/ GPU/  OpenCL& std::vector &RDFS entailment \cite{UJK2009} \\\hline
 Bre{\ss}  et.al.\cite{Bre2014} & GPU /CUDA & N/A & Optimization of workload of database operation \emph{aggregate, select,sort, and join}\\\hline
Groppe et.al.  \cite{Groppe201} & Cluster of computers & $B+$ tree&   Distributed merge join with indexing \\\hline
  Corcoglioniti et.al.  \cite{Corco} & Java/ Multithread &   HashMap &  RDF libraries   for building RDF processing pipeline\\\hline
  Our work & GPU/CUDA & TripleID &RDF query \emph{select, union, join, filter} and RDFS entailment \\\hline
\end{tabular}
}
\end{center}
\end{table}

%

 In this work, optimization of RDF storage utilizing  both CPUs and GPUs  was considered. The RDF data might  be stored and processed on GPUs or CPUs depending on the speed up dynamic measurement.
 Reasoning algorithms that are suitable for GPU computing were selected.  
The approach consists of three steps: optimizing SPARQL aggregate and ordering using CUDA reduction,
parallel constraint check by GPUs, and dynamic materialization by the GPU.

 \subsubsection{Compressed  data formats}
Since  the GPU memory size is limited and the copying time to and from the GPU  memory  can degrade  overall performance, it is advisable  to compact data  before transferring.
 One of the pioneer efforts  on transforming and compressing the RDF representation was
   by Atre et al.  The representation was called BitMat, which
   stores relations in a  bit matrix: one matrix is  created for one predicate \cite{Atre2010}.
 Madduri and Wu  presented  a FastBit software tool using bitmaps compression \cite{MKK2011}. 
 Kim et al.  \cite{Kim2015} considered the binary Header-Dictionary-Triple (HDT)  \cite{FMPGPA13} form and processed RDF queries   using the GPUs. The bitmaps as  well as  dictionary in HDT were   loaded to the GPU memory.   The prefix sum was applied to compute  predicate and object positions in  bitmaps.  They experimented on a  simple set of queries.
HDT is a popular compressed format. However, the conversion to this form takes a lot of time and memory. 
For a  larger number of triples, HDT with Java interface was required to increase Java heap memory to  handle more elements in  the set  and  would take even more  conversion time or C implementation should be applied.
 The  paper, however, did not address the issue of speeding up  the conversion process and data scaling.
The bitmap itself is compact   in a storage but when queried, bitmaps, dictionary information  are needed.  Such information must also be loaded into the GPU memory for searching and the conversion from such data structure to suit the GPU memory layout is required.

 For a very large RDF file,  Hexastore with MPI was used to support a cluster processing \cite{hexastore}.   Hexastore data can be split across the nodes in the cluster so that a concurrent query can be performed.
Thus, a file splitting  is another approach to handle concurrent searches. Simple file splitting scripts take a lot of time to run, hence using MapReduce to process large files is another possibility to run  on a cluster which is recognized as batch processing. 
We may consider  stream processing in the future with overlapping the memory transfer and the computation.
Interesting   Merge-Join operation in the GPU library introduced by Baxter \cite{Sean2013}   is generic 
 based on unified memory, and easy to use. 
 Often,  the  number of merged results may be too large; thus, on a GPU host computer   whose  memory size equals 12G, it was possible to apply the libraries when the number of elements of each vector was  around  5-6 thousands.  

In this work, we consider processing  the RDF data. We begin  with considering   the traditional search algorithm. The search algorithm can be customized in the framework. 
 Although it is possible to use a fast search, the fast search usually needs preprocessing such as creating prefix/suffix tables or implicit state machines. The construction of the preprocessed table requires  space and time overhead for different search strings \cite{chantana2}. 
Our framework assumes simplicity by using thousand threads to do  brute force matching.   From the preliminary study and previous work \cite{ Kouz},   with thousand threads, the gained  speedup with the optimized search scheme may not be significant  considering preprocessing overhead.
 
The framework  transforms  the RDF data into the TripleID format which encodes IRI  strings into IDs.
 The TripleID data   are then transferred into the  GPU memory.  
After that,  concurrent threads search the required triples according to the given query.  Indexing scheme is  currently not considered. It  is possible to create an  index  based on a tree structure such as HDT.   The concurrent search scheme is also possible with indexing e.g. by subjects \cite{Kim2015}.  Note that the GPU memory  is also required to store index information for each tree level. With more indexing types, more memory space is needed.

In the following section, the framework is first presented and the algorithms for different query operations are described based on   TripleID.

\section{TripleID-Q: Processing Framework} 
\label{sec:method}
The challenges of this research are to process the big data set with the limited GPU memory and  to simplify the representation  properly for GPU computation. 
The design goal is as  follows. 1) The format should be
simple  so  as to minimize  the conversion overhead.  
2)  It should not occupy  too large space.
3) Since the GPU has a lot of threads to help search, we will not focus on the index construction,  rather we intend to use  the large number of threads to look for  the data.

\begin{figure}[!t]
\centering
\includegraphics[width=3.6in]{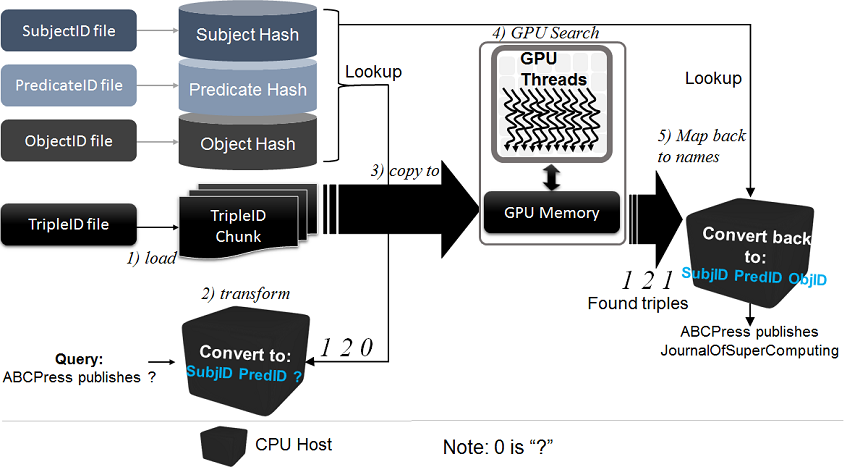}
\caption{TripleID-QE: Overall process.}
\label{fig:framework}
\end{figure}

TripleID-Q framework contains components to perform input conversion, look for query answers, and return the results
based on such a representation,  presented in  Figure \ref{fig:framework}. The RDF file (N3/N-Triple type) is transformed into four files as, Subject ID, Predicate ID, Object ID and TripleID files.The first three files are in the same format containing tuples in a form: \texttt{(keyID, value)},  where \texttt{key} is an integer and \texttt{value} is a string.
The TripleID file contains only triples in the form \texttt{(SubjID, PredID, ObjID)} and is a binary file assuming each 32-bit unique ID.   When loading these ID files in to memory, \emph{zlib} \cite{zlib} may be used to encode the values to save memory space for text. In theory, the size of IDs is $\max (\lg n_1, \lg n_2, \lg n_3)$ where $n_1$ is the total unique term used for subjects, $n_2$ for predicates, and $n_3$ for objects respectively.

In Figure \ref{fig:framework},  Subject ID, Predicate ID and Object ID ID files are loaded into memory in Step  (1).  
We use hash tables to store the tuples,  \texttt{(key, value)} pairs.
The given query is transformed into a  triple form \texttt{(? P ?)} (2), where \texttt{P} is the predicate ID.
For example, to search \texttt{ABCPress publishes} which journals, in Step 2,
\texttt{ABCPress} and \texttt{publishes} are transformed into  \texttt{SubjID, PredID}, which are
\texttt{1,2}, respectively.
The query becomes \texttt{1,2,?}.
In Step 3, the Triple ID file is split into chunks and the chunk is loaded into GPU memory.
Then, GPU threads concurrently look for  \texttt{1,2}, in the GPU memory. The found triples are marked and  returned.
In Step 5, the TripleID \texttt{1,2,1} is mapped back to the values
using the hash tables.

\begin{figure}[!t]
\centering
\end{figure}

\begin{sloppypar}
The framework is described as shown in  Algorithm \ref{al:algorithm1}.
A TripleID file is read by chunks.  It is assumed  that the keys to search are in array  \texttt{key.subj, key.pred, key.obj}, corresponding to Subject ID, Predicate ID, and Object ID respectively, where value 0 is reserved to represent a free variable "\texttt{?}".
For each thread, the kernel code,  $GPUSearch$  is executed.  $GPUSearch$ 
depends on a  selected search algorithm. This work implements a brute-force matching which 
finds the matches between given \texttt{key.subj,key.pred,key.obj}, corresponding to Subject ID, Predicate ID, and Object ID, accordingly.

 \begin{algorithm}[!htb]{\small{\setstretch{0.7}
\caption{ Parallel Search for TripleID }\label{al:algorithm1}
\KwIn{$\textit{dataArray}$, $\textit{key}$}
\KwOut{$\textit{positionArray}$}

 Allocate device memory for $\textit{dataArray}$, $\textit{key}$, $\textit{positionArray}$. \\
 \While {$not$  $EOF$} {
    Read a TripleID  chunk   in $\textit{dataArray}$.\\ 
Copy $\textit{dataArray}$, $\textit{positionArray}$ (initialized to false) and copy \textit{key} to the GPU memory \\
Call $GPUSearch$   with $dataArray$, $key$, and $positionArray$\\
 Copy $\textit{positionArray}$ back to the host.
 Map $\textit{positionArray}$ to corresponding  triples found. 
}
Free all the memory.
}}\end{algorithm}

A TripleID chunk is stored as \texttt{dataArray}  in the GPU main memory. 
Thread $i$ compares  \texttt{dataArray[i],dataArray[i+1],dataArray[i+2] } to  \texttt{key.subj,key.pred,key.obj}, corresponding to Subject ID, Predicate ID, and Object ID).  \texttt{DataArray} has total size $N$, $positionArray$ has a size, $N/3$, since the only found triple positions  $i$ are marked.

\end{sloppypar}

\begin{figure}[!htpb]
\centering
\includegraphics[height=1.3in]{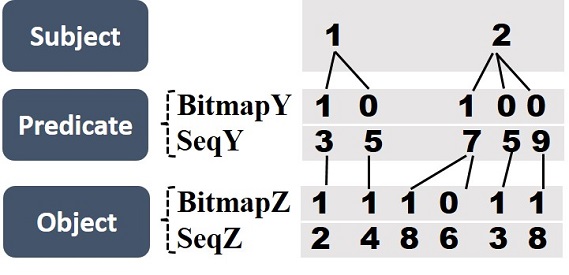}
\caption{HDT representation.}
\label{fig:hdt}
\end{figure}

 \begin{sloppypar}
Each element of  \texttt{dataArray} is  \texttt{ID} type.
The \texttt{key}contains three elements of \texttt{ID} (which is 32-bit each). 
Total memory size used by all these arrays in GPU memory is  $(N+\frac{N}{3}+3)\times $\texttt{sizeof (ID)}, where 
$N$ is the size of \texttt{dataArray}. For the \texttt{positionArray}, there may be other possible implementation such as keeping  the found positions in a list
and use \texttt{atomic} operation to eliminate the race condition in updating a list of found positions.
Compared to HDT representation in Figure \ref{fig:hdt}, the total memory size required is
\texttt{sizeof(BitmapY)+\thinspace sizeof(SeqY)} \texttt{+\thinspace sizeof(BitmapZ)+sizeof(SeqZ)}, which is $2\times$\texttt{sizeof(IDs)}\thinspace + \thinspace $2 \thinspace \times$\texttt{sizeof(Bitmap)}. This is just for querying in the order of subject, predicate, object respectively.

\end{sloppypar}

$GPUSearch$ can  also be  modified to accommodate  indexed triples.
 Though using the indexing scheme can make the search fast, it requires
  preprocessing time for   index information   and  requires more memory space and data  transfer for keeping indices in GPU memory.
For example, consider storing as  the HDT representation,  which is indexed by subjects.
HDT contains a collection of trees as depicted  in Figure \ref{fig:hdt}.
The first level contains all subjects where subject IDs are implicit, i.e., in an increasing sequence of 1,2,3 $\ldots N$, where $N$ is a total number of distinct subjects.
In the second and third levels,  $ SeqY$ and  $SeqZ$ are  lists of PredID and ObjID.
$BitmapY$ and $BitmapZ$ are  markings of  starting positions for predicates and objects respectively.
Thus, all of the four  arrays must be transferred to the GPU memory,
and the concurrent search must be done through   $BitmapY, BitmapZ,  SeqY$, and $ SeqZ$ \cite{panu}. Also, only the thread numbers  that are related to indices 
  performs the search.
  Compared to this work,  we generally consider storing TripleIDs without any index   and use lots of threads to directly search through them. The preprocessing requires only for the data conversion, but the indexing process is not required. 

Based on Algorithm \ref{al:algorithm1}, it is easy  to handle multi-GPUs and a cluster of GPUs whenever more host memory is  available  in Line 3 of  Algorithm \ref{al:algorithm1}, we can read each chunk for each GPU and in Line 6, the search kernel is called for each GPU. The results are aggregated from all GPUs and may be exchanged between GPU memory.
   CUDA-aware\cite{cudaaware} can be setup to combine \texttt{MPI\_Send} and \texttt{cudaMemcpy} together in one command and  chunks are distributed to each node.
\section{Handling Multiple Query Operations}
\label{sec:method2}
Previous section demonstrates a mechanism to handle a single query for triples where  the query for subject, predicate, object and any two combinations are possible.  
In this section, we  explain the handling of union and join operations of subqueries (also called \emph{triple patterns}).

\subsection{Union Operation}
\label{sec:union}
One query contains many subqueries, for instance, the query from \cite{Sai2014}:

\begin{lstlisting}[style=mymatstyle, 
language=SPARQL,
showspaces=false,
basicstyle=\ttfamily,
commentstyle=\color{gray},
caption = {Query with union 1},
label = listing3]
SELECT * WHERE {
 {<http://dbpedia.org/resource/Cabezamesada> 
     rdfs:comment ?var0 . }
UNION{<http://dbpedia.org/resource/Cabezamesada> 
     foaf:depiction ?var1 .} 
UNION{<http://dbpedia.org/resource/Cabezamesada>
     foaf:homepage ?var2 .}}
\end{lstlisting}
%

The above query consists of three subqueries of the same triple pattern \texttt{S P ?}. 
In this example, each triple can be a result of only one triple pattern. 
 

However, for the query as following, there are two variables in each subquery, where a triple may be the answer of  two subqueries.
For example, the  triples that are answers of the first subquery  
may also be the answer of the second subquery, \texttt{?var2}  \texttt{foaf:depiction}  \texttt{?var3}. That is
\texttt{?var2} may be    \texttt{<http://dbpedia.org/resource/Cabezamesada>},
 \texttt{foaf:depiction}  may be \texttt{?var0} and \texttt{?var1} may be  \texttt{?var3}. 

\begin{lstlisting}[style=mymatstyle, 
language=SPARQL,
showspaces=false,
basicstyle=\ttfamily,
commentstyle=\color{gray},
caption = {Query with union 2},
label = listing4]
SELECT * WHERE {
    { <http://dbpedia.org/resource/Cabezamesada>  ?var0 ?var1 . }
UNION {?var2 foaf:depiction ?var3 . } } 
\end{lstlisting}


 Previous implementation in Section 3 assumes  that a triple can be the answer of only one subquery. 
For a query containing more than one subqueries, it is required to indicate  that the triple is  the answer of which subquery of a further join operation.
Thus,  the data structures are modified as follows:
\begin{enumerate}
\item   \emph{positionArray} element is expanded to contain an array of subqueries, i.e.,
\emph{positionArray[i].query} contains the list of  subqueries where   triple $i$ is  served as an answer.
For instance, \emph{positionArray[i].query }= \{1,3\}, implies that triple  $i$ is an answer for subqueries 1,3. 
\emph{query} can also be implemented as a fixed-size array since 
typically, a query contains a small constant number of subqueries which is a small amount (1-10).
This amount can actually be analyzed in the preprocessing phase during the query parsing. 

\item  \emph{key} array  is enlarged to the size of multiples of threes to hold TripleIDs for many subqueries.  
\end{enumerate}

\begin{figure}[!thpb]
\centering
\includegraphics[width=0.8\columnwidth]{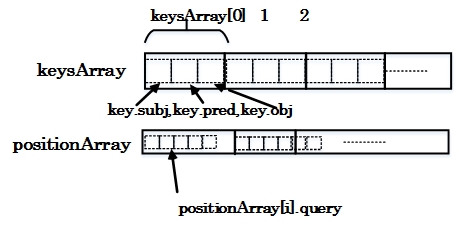}
\caption{Modification of  \emph{keysArray, positionArray}.}
\label{fig:mod}
\end{figure}

Variable \emph{key}   becomes \emph{keysArray} and the loop performs
for every element in   \emph{keysArray}.   \emph{ positionArray}  
becomes \emph{positionArray[i].query}, which is to mark triple $i$ matched for subquery in   \emph{query}  in Figure \ref{fig:mod}.

\subsection{Join Operation}
\label{sec:join}

Results from several subqueries can be joined regarding the  relation between each subquery. Example in Listing 5  \cite{Sai2014} contains 5 subqueries, each of which is the pattern according to Table \ref{tab:listing5}.

\begin{lstlisting}[style=mymatstyle, 
language=SPARQL,
showspaces=false,
basicstyle=\ttfamily,
commentstyle=\color{gray},
caption = {Query with 5 subqueries},
label = listing5]
PREFIX rdf:<http://www.w3.org/1999/02/22-rdf-syntax-ns#>
PREFIX ub: 
<http://www.lehigh.edu/~zhp2/2004/
0401/univ-bench.owl#>
SELECT ?X, ?Y1, ?Y2, ?Y3
WHERE {
  ?X rdf:type ub:Professor .
  ?X ub:worksFor <http://www.Depart0.University0.edu> .
  ?X ub:name ?Y1 .
  ?X ub:emailAddress ?Y2 .
  ?X ub:telephone ?Y3 .
}
\end{lstlisting}

  \begin{table}[thbp]\small
\caption{Subquery pattern of Listing  5}
\label{tab:listing5} 
\centering
\setlength{\tabcolsep}{0.5pt}
\def\arraystretch{0.8}
 
\resizebox{\columnwidth}{!}{%
\begin{tabular}{|ll|c| }\hline
&Sub-query & Triple \\
&&pattern\\\hline
 $q_0$:&\texttt{?X rdf:type ub:Professor} & \texttt{\{? P O \}}\\
  $q_1$:& \texttt{?X ub:worksFor }&\\
       &\texttt{<http://www.Depart0.University0.edu>} & \texttt{\{? P O \}}\\
  $q_2$:&\texttt{?X ub:name ?Y1} &  \texttt{\{? P ? \}}\\
  $q_3$:&  \texttt{?X ub:emailAddress ?Y2} &  \texttt{\{? P ? \}}\\
   $q_4$:& \texttt{?X ub:telephone ?Y3} &  \texttt{\{? P ? \}}\\
   \hline
\end{tabular}
}
\end{table}



The relational join implementation is based on the design pattern of CUDA library by \emph{Modern GPU}  or $Mgpu$  \cite{Sean2013}.
Specially, we use the  function \emph{RelationalJoin<MgpuJoinKindxxx>}, where xxx can be either Inner, Outer, Left, Right  types of  join. 
  Before calling such function, the data needed to be sorted and merged.   \emph{Mergesort} and
\emph{Merge} in  \emph{Mgpu} library  are also used. The steps  to incorporate the join  operation to the proposed framework are summarized  as follows.

\begin{enumerate}

\item Analyze the relation between subqueries. There are at most 9 possible types of relations between any two subqueries.
Without loss of generality, assume  the subqueries are ordered based on the original query.
For two subqueries $q_i$ and $q_j$, where $i < j$, $q_i$ may be related to $q_j$ in one of the following relationship types  \{OO, PP, SS, OP, OS, PS, PO, SP, SO\}.
 Relationship OO implies that $q_i$ is related to $q_j$  using O (as the object is the same for both $q_i$ and $q_j$). 
 
Table \ref{tab:reldef}
presents the definitions and examples of these relationships.
As an example, consider Row  OP. The objects of subquery $q_i$ is related to predicate of subquery $q_j$.

%

In Table \ref{tab:listing5}, subquery $q_0$, is related to $q_1$   by variable ?X; thus, its relationship is    type SS.
The  relationships of these subqueries  are $REL=$  \{\{$q_0$,$q_1$,SS\}, \{$q_1$,$q_2$, SS\},  \{$q_2$,$q_3$, SS\}, \{$q_3$,$q_4$, SS\}\}.
That is $q_0$ is related to $q_1$ as SS, $q_1$ is related to $q_2$ as SS, and etc.

\item Perform the execution of each subquery as in Algorithm \ref{al:algorithm1}.
To  apply the merge-join library  using $Mgpu$,  the triple results are re-organized as vectors shown in Figure \ref{fig:vector}.  

Vector $R_{q_i}$  stores triple results from  subquery $q_i$.  Each vector $i$ has length $n_i$. Such vector contains two parts (key, value). The key and value parts depend on the type of relations. For example, if two subqueries  are related as SS,i.e., \{$?S$, $P_1$, $O_2$\},  and \{$?S$, $P_2$, $O_2$\}, the key of both vectors are subjects $?S$ and the values are the remainder parts.

 \begin{table}
\centering
\small
\caption{Relationship between two triple patterns.} 
\label{tab:reldef} 
\begin{tabular}{|c||c|}\hline
Relationship  & Example related triple patterns \\
types  & of subqueries $q_i$ and $q_j$\\ \hline
OO & \{$S_1$, $P_1$, $?O$\},  \{$S_2$, $P_2$, $?O$ \}\\
PP &\{$S_1$, $?P$, $O_1$\},  \{$S_2$, $?P$, $O_2$ \}\\
SS & \{$?S$, $P_1$, $O_1$\},  \{$?S$, $P_2$, $O_2$ \}\\
OP & \{$S_1$, $P_1$, $?O$\},  \{$S_2$, $?O$, $O_2$ \}\\
OS & \{$S_1$, $P_1$, $?O$\},  \{$?O$, $P_2$, $O_2$ \}\\
PS & \{$S_1$, $?P$, $O_1$\},  \{$?P$, $P_2$, $O_2$ \}\\
PO & \{$S_1$, $?P$, $O_1$\},  \{$S_2$, $P_2$, $?P$ \}\\
SP & \{$?S$, $P_1$, $O_1$\},  \{$S_2$, $?S$, $O_2$ \}\\
SO & \{$?S$, $P_1$, $O_1$\},  \{$S_2$, $P_2$, $?S$ \}\\\hline
\end{tabular}
\end{table}

\item  Aggregate the proper key  for joining and submit the result vectors to 
\emph{RelationalJoin}, i.e., for each relation $r_k$ $\in REL$, for two relations that are related by two subqueries $q_i,q_j$. 
Assume its type of relation  as $T_{r_k} = SS$.
 Consider for two result vectors $R_{q_0}$ and $R_{q_1}$.
As in Figure \ref{fig:vector2}, (1) copy subject $S_{0,l}$, $l=0,..,n_0-1$  from result vector $R_{q_0}$ of subquery $q_0$,  and copy subject $S_{1,l}$, $l=0,..,n_1-1 $  from result vector $R_{q_1}$ of subquery $q_1$,  
 where $n_0$ is the total results of subquery $q_0$, and
 $n_1$ is the total results of  subquery $q_1$. Then, we add the subject results to the key  vectors.
We copy the remainder part (predicates and objects) and put them in the value 
 In (2), Merge-join in Mgpu is called with key vectors. The results obtained are indexed pairs that display the positions of both keys that are joined. The positions of both keys are used to extract  the corresponding value vector elements from the vectors in  (3).
  Similarly for the cases of OO, PP, OP, OS, PS, PO, SP, SO, the proper terms of the triples for each result from any  subquery $q_i$, $R_{q_i}$ and  results from subquery $q_j$, $R_{q_j}$   are copied as  keys for merging.
  The positions for pairs of  keys that match  are returned as a vector of an element index pair as depicted  in Figure \ref{fig:vector2}. 
 Then, the result vectors after joining are used  for the next join in the next relation in $REL$.  

 \begin{figure}[!thtp]
\centering
\includegraphics[width=3in]{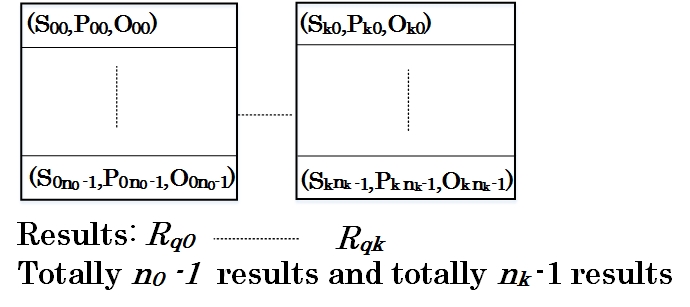}
\caption{Vectors of triple results.}
\label{fig:vector}
\end{figure}

 \begin{figure}[!thtp]
\centering
\includegraphics[width=\columnwidth]{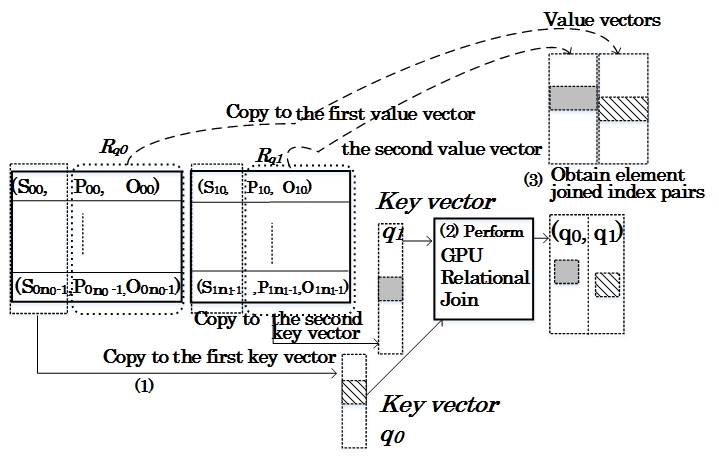}
\caption{Processing  join operation for SS.}
\label{fig:vector2}
\end{figure}

\end{enumerate}

\subsection{Other Operations}

To handle other operations such as FILTER,
the query results are first obtained, then the IDs of terms must be converted back to string values. 
 A regular expression may be used to filter ID names of  the matched TripleIDs.

An extra structure is needed to only  keep  variables  in each subquery. 
The variables for each subquery are used to find relationships $REL$ are discussed in  Subsection \ref{sec:join}. 
From a SELECT statement, the selected variables must be returned.
To handle DISTINCT, a hash table  is used to store the results of a variable.  
Various GPU  hash table versions are suggested in the literature \cite{Alc2011,cudpp}.
In the future, finding a good ID assignment of  subjects, predicates, and objects in such a way to preserve the ordering
and to filter out part of triples that are not relevant to the subquery is  an interesting problem. The total remaining triples will reduce the size of the GPU memory used in performing operations such join, union, etc.
 
 \begin{figure}[!t]
\centering
\includegraphics[width=\columnwidth]{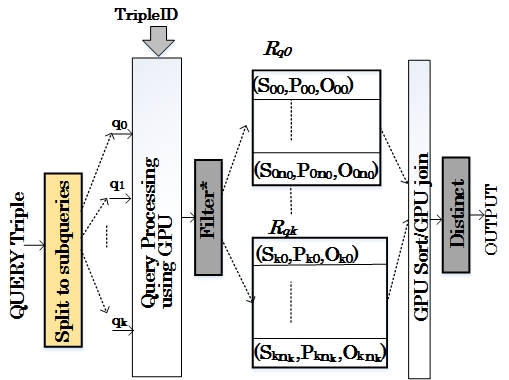}
\caption{Integration of several operations.}
\label{fig:vector3}
\end{figure}

Figure \ref{fig:vector3} presents  an overall process when mixing these operations, i.e. SELECT, DISTINCT, JOIN. After the query is split into subqueries $q_0, ... ,q_k$, 
each subquery is searched against  the TripleIDs by GPU threads. The resulting triples are marked as the answer of a subquery and the marked triples are extracted to store in the vectors corresponding to the subquery. The filter is used during this step.
The join operation starts from the left result $R_{q_0}$ to the right one $R_{q_k}$. Note that before joining, each result vector must be sorted. After joining all results, the final results are merged to keep only distinct values. When considering query optimization, join ordering can be changed.

\section{Experiments}
\label{sec:exp}
 The experiments demonstrate the  efficiency of the framework  in the following aspects.
First,  the  conversion time to TripleID format is compared to the conversion to other formats and
 the size of TripleID file is    compared to the original file type such as RDF and N-Triple file, and other formats such as HDT and RDF store.
Next, the search time to the these files is measured in various aspects: the number of subqueries,
the number of input triples, and different operations.

%
 
The tested machine  had  the following specification: 
  Intel$(R)$ Core$(TM)$ $i7-5820K$ CPU $@$ 3.30GHz, 6 cores, and 16 GB RAM with 
  NVIDIA Tesla K40.  The  card contained 15 Multiprocessors, 192 CUDA cores per MP (totally 2,880 CUDA cores)
with maximum clock rate 745 MHz (0.75 GHz).  Memory bus width was 384-bit. Total amount of global memory was 12GB.
 The targeted thread block size and grid size equal to 1024  and   480 respectively, which yield the best performance on our machine. Other tests that explore the other block size and grid size are demonstrated in \cite{chantana2}.

\subsection{ Data Sets}

\label{sec:redland1}

 Two data sets  are  considered: Billion Triples Challenge Data Sets  and SP$^2Bench$ Data Sets.
 The first data set was obtained from
 Billion Triples Challenge \cite{btc-2009}.
The downloaded contents 
  encoded in N-Quads format \cite{NQUAD}  were split into chunks of 10 million ($10^7$) statements, called chunk 01,02,..,07,  each of which has a size of 350 MB.
These splits were combined to obtain the files with  various sizes as shown in \ref{tab:dataset}.  
A  whole   crawled data   available as  ``BTC-small'' has size    equal to 2.172 GB.
These files were converted into an N-Triple format \cite{ntriple} format
 The conversion program (command-line tool), \texttt{rdf-convert-0.4} (http://sourceforge.net/projects/rdfconvert/), was used.


\begin{table}[thpb]\small
\caption{Data set characteristics (BTC)}
\label{tab:dataset} 
\centering
\setlength{\tabcolsep}{0.5pt}
\def\arraystretch{0.8}
 
\addtolength{\tabcolsep}{0.5pt}  
\begin{tabular}{|  l ||r|r|r||r|}\hline 
 data set& \# subj	&\#pred	&\#obj	&\#triples\\\hline
01	& 314,285 &	 3,458 	& 583,555 	& 1,868,651 \\
0103&	 778,772 	 &5,849 	& 1,383,943 	& 5,160,648 \\
0203	&  504,082& 	 4,477 &	 990,414 	 &3,291,997 \\
0207	&  366,654 	& 3,563 	& 688,019 &	 2,017,469 \\
012347&	 1,113,824 &	 7,542 	& 1,674,407 	& 7,083,790 \\
BTC-small&	 1,383,542 	& 8,205 	 &2,260,819 	& 9,627,877 \\\hline
\end{tabular}

\end{table}

For  SP$^2Bench$ \cite{vir2010},  the data sets were generated with different number of triples 
  up to 100 million triples.  SP$^2Bench$ produces the data sets in an N3 format \cite{N3}.
These files contain various numbers of subjects, predicates
and objects as shown in 
 Table~\ref{tab:dataset2}. 
 
 \begin{table}[thpb]\small
\caption{Data set characteristics (SP$^2Bench$).}
\label{tab:dataset2} 
\centering
\setlength{\tabcolsep}{0.5pt}
\def\arraystretch{0.8}
 
\addtolength{\tabcolsep}{0.5pt}  
\begin{tabular}{|l||r|r|r|r|}\hline 
 data set (triples) & \# subj	&\#pred	&\#obj	&\#triples\\\hline
5M		&896,359	&76	&2,400,922	& 5,000,120 \\
10M		&1,712,642	&77	&4,662,411	 &10,000,091 \\
20M		&3,404,855	&153	&9,379,299	 &20,000,429 \\
50M		&8,639,994	&306	&24,058,862&	 50,000,100 \\
100M	&	17,652,609	&613	&48,965,319	& 100,000,144 \\\hline
\end{tabular}
\end{table}

\subsection{Tools' Description}
Our following experiments show the various tested tools.
 The gathered tools focus on RDF querying with free, open source development: Redland, Menthok, Stardog, Virtuoso, and HDT.
 They have various implementations.  HDT has both  C and Java implementation and interfacing.
 In the experiments,  C implementation  is used  for Redland, Mentok and HDT.  
 Implementation for HDT has an indexed supported for $SPO$. Virtuoso is the largest one with an open source support for large RDF data while Stardog community edition can support around 20 million triples while   larger RDF data is supported with the enterprise version and free for trial for 30 days.

\subsection{Preprocessing Time}
We measure the preprocessing of using different formats.
 The conversion to TripleID  time is investigated and compared to the conversion time to HDT from the original NT format.
Then we measure the loading time, the case of using  these RDF stores, which  reads and parses RDF files (and construct an internal graph model in some cases).

\begin{table}[thpb]\small
\caption{ Loading time  in seconds using Redland, Mentok, and  TripleID  representations on $BTC$. }
\label{tab:compare-load-btc} 
\centering
\begin{tabular}{|l||r|r|r|}\hline
data set & Redland&  Mentok& TripleID \\ \hline

01 &	14.89	&111.87 	&0.52 \\
0103	&46.84&	261.58 &	1.33 \\
0203	&31.66&	166.84	&0.92 \\
0207	&16.464	&106.23&	0.66 \\
012347	&68.64&	369.90 	&1.95 \\
btc-2009-small&	83.77&	N/A&	2.4 \\\hline

\end{tabular}
\end{table}

Table \ref{tab:compare-load-btc} presents the loading time for each tool for the data set in Table \ref{tab:dataset}.
 Redland library consumes more time
to load the RDF file and construct the  graph model.
Note that the query time of Redland  is about 
1/2 or 1/3 of the model loading time.
It is found that Mentok's loading time was much more that of Redland while the query processing could obtain benefits from multiple MPI nodes.
From this observation, when the number of triples becomes very large,  the straight-forward program which reads   RDF triples and creates a
  simple representation will save this preprocessing overhead.


\begin{table}[htbp]\small
\caption{ Loading time  in seconds for SP$^2Bench$ using Stardog, HDT, TripleID.}
\label{tab:compare-load-sp2}
\centering
\begin{tabular}{|l||r|r|r|}\hline
data set & Stardog &  HDT& TripleID \\ \hline
5M	&40.98&	0&	1.86 \\ 
10M	&873.98&	0&	4.1 \\
20M	&3,820.71	&0.01	&8.66\\
50M	&424.54&	0.03	&19.65\\
100M	&1171.36&	0.05&	42.56\\\hline
\end{tabular}
\end{table}

\begin{table}[tbhp]\small
\caption{ Comparison for conversion time  (HDT and TripleID) in seconds for $BTC$ }
\label{tab:compare-conv-btc} 
\centering
\begin{tabular}{|l||r|r|r|}\hline
data set & HDT	&TripleID	&Speedup HDT/ \\ 
& (s) & (s) &TripleID\\\hline
01		&	19&	3.25	&5.85\\
0103	&		51&	8.57&	5.95\\
0203	&		34	&6.15&	5.53\\
0207	&		22	&3.06&	7.19\\
012347	&		71	&10.37&	6.84\\
btc-2009 &			94&	28.86	&3.26
\\\hline
\end{tabular}
\end{table}

\begin{table}[thbp]\small
\caption{ Comparison for conversion time  (HDT and TripleID) in seconds for $SP^2$ }
\label{tab:compare-conv-sp2} 
\centering
\begin{tabular}{|l||r|r|r|}\hline
data set & HDT	&TripleID	&Speedup HDT/ \\ 
& (s) & (s) &TripleID\\\hline
5M		&	56	&14.37	&3.90\\
10M		&	62	&31.04	&2.00\\
20M		&	231	&62.08	&3.72\\
50M		&	360	&148.44	&2.43\\
100M	&		1256&	298.1	&4.21
\\\hline
\end{tabular}
\end{table}

Table \ref{tab:compare-load-sp2} compares the loading time of SP$^2Bench$ in  Table \ref{tab:dataset2}.  SP$^2Bench$ generates larger number of triples.  We compare the loading time of RDF data using the large triple store, Stardog \cite{stardog}. To support large number of triples, the setting of Stardog was -- 
JVM memory is 8G and
Off heap memory is 64G.
 Stardog prefers the triples to be in the Turtle format or called in short, TTL  \cite{turtle}.
 Thus, RDF data were converted into TTL format.
The reported time under "Stardog" column
 is the time used to load TTL files into Stardog data store\footnote{  The Stardog's   reported time for loading in total and in Triples per second. This 20M case has a slowest is around 4.4K triples/sec while for other cases, 5M is 113.5K triples/sec,  10M is 11.4K triples/sec, 50M is 103.5K triples/sec, and 100M is 83.1K triples/sec.}.
Under "HDT" column, the time for loading HDT data is shown.
The time to load the HDT data set is very small compared to others
since the HDT file is already small.

Table \ref{tab:compare-conv-btc} and Table \ref{tab:compare-conv-btc} display the conversion time to the TripleID format compared to the conversion time to   the HDT format for $BTC$ and SP$^2Bench$ respectively.  HDT with C implementation (\emph{rdf2hdt}) is used for comparison. 
Conversion time to  the HDT format is  about 5 times   longer than that of TripleID files for $BTC$ and about 2-3 times longer for  $SP^2$.


\subsection{Compaction}
\begin{figure}[!thbp]
\centering
\includegraphics[width=3.5in]{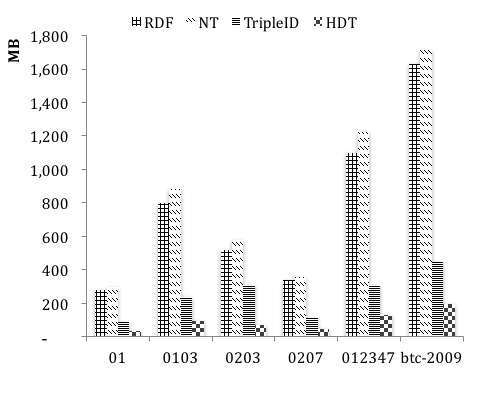}
\caption{Comparison of data size   for BTC. }
\label{fig:compare1}
\end{figure}

\begin{figure}[!hbpt]
\centering
\includegraphics[width=3.5in]{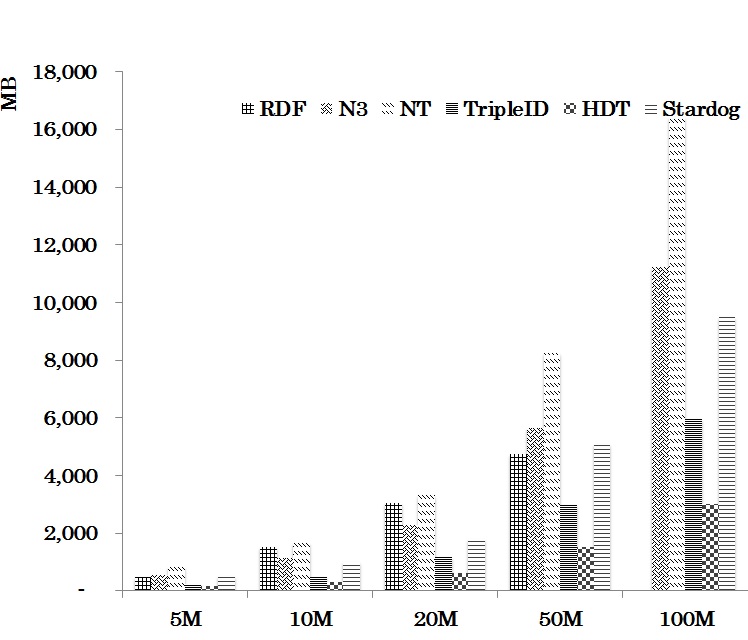}
\caption{Comparison of data size for SP$^2Bench$. }
\label{fig:compare2}
\end{figure}

Figure \ref{fig:compare1} and Figure \ref{fig:compare2} compare the file sizes after the conversion to TripleID for BTC and  SP$^2Bench$ respectively.

As in the previous section, 
after transforming to TripleID format, the four files are generated.
The file size in Column "TripleID" is the summation of TripleID file size plus the subject, predicate, object ID files' size. The sizes are compared against the original RDF,  NT  and HDT files.
TripleID   size compared to NT size is around 3-4 times smaller. However, TripleID size is 2 times larger than that of HDT format   since we do not eliminate redundancy (due to shared subject and object elements) and we do not perform the compression while as noted in the above subsection, the TripleID conversion time is about 3 times faster than the HDT conversion time. We also tried convert some large data set such as `012347'  and `btc-2009' to Stardog format and we found that the size of Stardog format is around 1/2 of that of NT format. 
For  the large case, SP$^2Bench$, we compare against $N3$ and Stardog. $N3$ is smaller than NT size and our TripleID size is smaller than that of Stardog database.


\subsection{Single Subquery Speedup}
Table \ref{tab:compareredland} displays  processing time of each simple query containing one subquery.

  {\singlespace\small
\begin{lstlisting}[
           language=SQL,
           showspaces=false,
           showstringspaces=false,
           basicstyle=\ttfamily,
           numbers=left,
           numberstyle=\tiny,
           commentstyle=\color{gray},
        ]
 SELECT distinct ?subject ?object 
WHERE { ?subject owl:sameAs ?object} }.
\end{lstlisting}}

Column "Redland" shows the query time using Redland. The Redland library for this test was modified so that it can handle larger models. 
Using the traditional Redland library to search reaches the memory heap limit for allocation of a graph model storage whose size was larger than that of 01 case (1.8M triples), due to
the growth of the internal  model, represented by the hash table.  Redland library reallocates the model whose size is   double  to the current one when the hash table density is more than 50\%. The machine  could not allocate large continuous heap memory  area to store   the  model, which made the program stops running.
We, then, modified Redland source code to split into  smaller submodels   and to link  the submodels as a list iterator.
The splitting was done after parsing of the input RDF file by Rasqal parser.  

Column "Mentok" shows query time using  Mentok which is the reimplementation of Hexastore \cite{hexastore} and the addition of   MPI \cite{mentok}.  This one demonstrates the use of distributed RDF models. Testing this library, 
we deployed Mentok on a cluster of  4 nodes with  MPI, where each node was  Intel(R) Xeon(R) CPU   X3470  @ 2.93GHz.  Column "HDT" displays the query time using HDT library (C implementation)\cite{FMPGPA13}. These reported numbers are  query time excluding loading time.
Column "TripleID" is our search time.
The speedup for each case (TripleID over Redland, TripleID over Mentok and TripleID over HDT) is displayed under column "Speedup".

\begin{table}[thbp]\small
\caption{Time comparison in  seconds  between Redland, Mentok, HDT and TripleID for a simple query.}
\label{tab:compareredland} 
\centering
\setlength{\tabcolsep}{0.2pt}
\def\arraystretch{0.6}
 
\addtolength{\tabcolsep}{0.5pt}  
   
\begin{tabular}{|l||r|r|r|r|r|r|r|r|r|r|}\hline
& & & & &
  \multicolumn{3}{c|}{Speedup}\\\cline{6-8}
 data set & \multicolumn{1}{c|}{ Redland }& \multicolumn{1}{c|}{Mentok}& \multicolumn{1}{c|}{HDT}&\multicolumn{1}{c|}{TripleID }&
\multicolumn{1}{c|}{\underline{Redland}}& \multicolumn{1}{c|}{ \underline{Mentok}}& \multicolumn{1}{c|}{ \underline{HDT}}\\ 
    & 	 & &	   &  	&TripleID& TripleID& TripleID\\\hline
01	& 	6.29	&   0.59	  &0.16	&0.13	& 48.38	&4.57 &	1.23\\
0103&	 	28.22& 1.82&0.43&0.20& 141.10	&7.26&	2.15\\
0203	& 	21.31 & 1.16& 	 0.37&	0.23	&92.65&	5.06&	1.61\\
0207	& 11.55& 0.70& 0.19&0.12&  96.25&	5.86	&1.58\\
012347&	 	36.98&	 1.90& 	 0.69&0.18	& 205.44&	10.55&	3.83\\
BTC-small&	 35.39&   N/A  &0.79	&0.09	&393.27	&N/A&	8.78\\\hline
\end{tabular}
\end{table}

The speedup of  querying using TripleID over Redland  is significant which  is   about 48-390 times faster. 
Compared to the speedup  of querying over Mentok  for TripleID is   about 4-10 times faster.
 We could not  perform the test for BTC-small for Mentok since it used up the memory allowed in our cluster environment.
HDT gives a close performance to our TripleID when the number of triples are not very large but for a large data set the speedup is obvious (BTC-small). 
The speedup over HDT also depends on  the query types.
More speedup is gained when the number of triples are around 5 millions or more.
For 5-million triple data set (0103), the speedup is about 2 times and for 7-million triple data set (012347), the speedup is about 3-4 times.
Consider using RDFlib \cite{RDFlib}. On the same machine,   processing  5 million-triple data (with N3 size of 826,904,622 bytes)
took 778.22 seconds while we observed that the loading time  was 776.96 seconds and the query time was   1.25 seconds.
Thus, we could not  perform the larger test using  RDFlib since the process would use too much memory resource than allowed.

 Table \ref{tab:compare3}  shows the total query time for our TripleID form in Column "total time" for SP$^2Bench$.  This benchmark contains more number of triples.   we compare against Stardog 4.2.1, and HDT, with  
 the query  pattern   "?PO", .where  P is \texttt{rdf:type} and O is \texttt{foaf:Person}.
 The total time obtained from querying 100 million triples is 2.28 seconds where the triple conversion time was 298.1 seconds.


\begin{table}[thbp]\small
\caption{ Query  time  in seconds using TripleID, HDT, Stardog   on SP$^2Bench$. }
\label{tab:compare3} 
\centering
\begin{tabular}{|l||r|r|r|r|r|}\hline
data set & Stardog &  HDT& TripleID &\multicolumn{2}{c|}{Speedup}\\ \cline{5-6}
& & & & \multicolumn{1}{c|}{ \underline{Stardog}}& \multicolumn{1}{c|}{ \underline{HDT}}\\
& & & & TripleID& TripleID\\\hline
 5M	& 0.25 & 	1.21 &	 	0.25 &1.0	&4.8\\
10M	&	69.77& 1.96& 	0.65 & 107.3	&3.0\\
20M	& 78.04& 4.64	&0.77& 101.4	&6.0\\
50M	&424.54&  9.98	&1.66 & 255.7&	6.0\\
100M& 593.94& 22.38&		2.28 &260.5	&9.8\\\hline

\end{tabular}
\end{table}

\subsection{Multiple-Subquery  Speedup }

The performance of queries containing subqueries where each subquery contains union, join, or filter  is  measured.
 Particularly, the  selected data sets from previous subsections are considered, with the cases
of  5 million triples and 7 million triples,
namely 0103 and 012347  from Table \ref{tab:dataset}.

Three types of queries are considered with different focuses:  Q1-Q5 only  focus  on union operations,
Q6-Q8 focus  on filter and union operations, and
Q9- Q16 emphasize on join and filter operations. 
The join operation may be in the type of SS,  OS, or two  consecutive SSs or three consecutive SSs etc. 
Details of the queries are in  Appendix \ref{sec:appendix}.
Table \ref{tab:misc} presents  execution time  in seconds of our approach compared to that of  Redland, Virtuoso, Stardog, and HDT respectively.
In Column "\#Res", the number of final RDF results obtained for each query in RDF is shown, except in Q16, where the number shown is the number of NT triples. 
Under column "Redland", the load time and query time are presented.
Columns "Virtuoso" \footnote{ Virtuoso  
 7.2.2  \cite{virtuoso}   which is a column store as well as 
 \emph{isql} from OpenLink Interactive SQL (Virtuoso), version 0.9849b  were used. The default setting for Virtuoso was assumed.} \cite{virtuoso}, "Stardog" and HDT display the query time using these data stores respectively.
 Under column "TripleID", we display the time for loading TripleID, the time for transferring the data to GPU memory, the time for joining operations, and the time for querying, under "load", "data", "join" and "query" respectively.

Column "Speedup" shows the  query time speedup over   Redland, Virtuoso, Stardog, HDT respectively.
In most cases, using TripleID achieves  speedup  depending on query types.
For the union operations, as in Q2-Q4, the number of results is large compared to the results of Q1.
Also, in Q8 which contains three subqueries with  filter, and union operations, the computation time was increased compared to Q7 containing two 
 subqueries with union and filter operations.
The speedup is obvious when compared with Redland's query time which ranges 16-108 times.
 In the case of Q16 (N/A),  the query could not be executed using Redland because the query processing consumed all the memory resources and the execution was aborted.
The rows with "-" indicate that TripleID yields no speedup.
For Virtuoso, the speedup varies from 3-1,131 times. For Q5 or Q6, the query pattern is "S??"  where Virtuoso can perform very fast.    
Stardog performs queries much faster than Virtuoso for the queries that returns  large number of results.
 Stardog also performs well   when the the query pattern is "S??".
It gives fast join results for queries Q9--Q12. 
The union operation takes longer time.
  HDT running time \footnote{The queries for HDT were implemented using C.  Data structure used to store each subquery's results was  vectors.} is quite consistent and fast. 
The running time for HDT queries is very fast when the pattern is "S??" in Q5 and Q6. 
The speedup of TripleID over HDT is varied depends on queries. 
More speedup is obtained for Q2,Q3,Q4 for the large union results.
 
\begin{table*}[!tp]\small
 
\caption{Comparison between Redland and TripleID performance for BTC-0103 dataset. }
\label{tab:misc}
\begin{center}
\setlength{\tabcolsep}{0pt}

\begin{tabular}{|l|c|c|c|c|c|c|c|c|c|c|c|c|c|c|} \hline
Query & \#Res & \multicolumn{2}{c|}{Redland }    & Virtuoso&Stardog & HDT&  \multicolumn{4}{c|}{TripleID }     & \multicolumn{4}{c|}{\textbf{Speedup}}     \\ \cline{3-4}\cline{8-15}
               &   & load  & query       &      & &    & load & data & join  & query &    Redland  &    Virtuoso&  Stardog &  HDT \\\hline 
              
Q1           & 20,081    & 43.29 & 7.54    & 7.82 &2.38&  0.54  & 1.36 & 0.29 & -         & 0.36  & 20.94  & 21.72 & 6.61&1.50 \\\hline 
Q2            &784,648   & 43.38 & 40.1    & 670.69&509.90&    2.56   & 1.25 & 0.29 & -         & 0.83   & 48.31    & 808.06&614.33&3.08\\\hline 
Q3            &  870,890 & 43.09 & 57.04     & 785.38  &570.19&    3.37 & 1.15 & 0.29& -         & 0.86  & 66.33   & 897.67 &663.01&3.91\\\hline 
Q4           &891,102     & 43.11 & 72.39     &785.95    &596.51&  3.51 & 1.24 & 0.29 & -         & 0.83  & 82.36    & 897.67 &718.68&4.22\\\hline 
Q5          &   24  & 43.21 & 5.54    & 0.04      &0.14&   - & 1.15 & 0.27 & -         & 0.32  & 17.31  &-  &-&-\\\hline 
Q6          &   18  & 43.05 & 5.55     & 0.01    &0.27& -     & 1.15 & 0.29 & -         & 0.32  & 17.34   &- &-&- \\\hline 
Q7            &22    & 43.23 & 11.45    & 3.00     &0.15& 0.40   & 1.15 & 0.3  & -        & 0.36  & 31.81  & 8.34 &-&1.11 \\\hline 
Q8            &20,370   & 43.23 & 19.49     & 4.87  &1.49&  0.80   & 1.11 & 0.28 & -         & 0.4   & 48.73   & 12.17 &3.73&2.00 \\\hline 
Q9             &1  & 43.01 & 47.67      & 2.14     &0.14&  0.90 &0.9& 0.27 & 0.03      & 0.45  & 105.93  & 4.76 &-& 2.00\\\hline 
Q10          & 0    & 43.32 & 48.97       & 0.01   &0.18& 1.27 & 1.15 & 0.28 & 0.02      & 0.45  & 108.83  & -&-&2.82\\\hline 
Q11            &  98 & 43    & 5.61   &14.70     &0.17&   0.36  & 1.11 & 0.28 & 0*         & 0.34  & 16.50    & 43.24 &-&1.05 \\\hline 
Q12         &  1,529   & 43.08 & 6.17       & 16.49  &0.36& 0.46 & 1.11 & 0.27 & 0*         & 0.36  & 17.14 & 45.81  &-& 1.27\\\hline 
Q13          &  30,427   & 43.16 & 8.23       & 22.97   &2.62& 0.53  & 1.14 & 0.27 & 0.02      & 0.38  & 21.66  & 60.45 &6.89&1.39 \\\hline 
Q14         &144,845    & 43.09 & 15.36    & 58.53   &20.62&   0.98  & 1.11 & 0.28 & 0.3       & 0.68  & 22.59    & 86.08&30.32&1.44\\\hline 
Q15          & 5,595   & 43.97 & 6.82       & 1.28    &0.65&   0.43& 1.29 & 0.27 & 0.03      & 0.42  & 16.24  & 3.05 &1.54&1.02 \\\hline 
Q16          &86,824    & N/A     & N/A    & 1,528.03    &614.34&  9.86      & 1.11 & 0.3  & 0.95      & 1.35  & -  & 1,131.88&455.07&7.30 \\\hline 
\end{tabular}
\end{center}
\end{table*}


Table \ref{tab:misc2} presents  timing results for the larger BTC data set (012347) containing 7 million triples.
The speedup trend is shown in the similar manner as in Table \ref{tab:misc}.  More speedup is gained  when compared to  Redland, Virtuoso, Stardog and HDT, especially in Q1,Q2,Q3,Q4,Q14,Q16.
The results imply that  the total execution time depends on the query operations and the number of results of  the certain query.

In some case,  
 the number of final results does not   reflect the total time since it also depends on the number of intermediate results
before joining.
The time in Column "join", indicated by 0*, which is closed to zero in Q9-Q12, 
implying that the number of intermediate results are small.
When the join time such as in Q14  is detectable, the number of intermediate results  is significant. 
In Q14, the first, second, and third subqueries return 22,626 results. 
In Q15,  the first subquery returns 22,626 results and the second subquery returns  6,300 results.
For the join with large intermediate results, the speedup will be more.


\begin{table*}[!thp]\small
\centering
\caption{Comparison between Redland, Virtuoso, Stardog, HDT, TripleID performance for btc-012347 dataset (time in seconds). }
\label{tab:misc2}
\setlength{\tabcolsep}{0pt}  
\begin{tabular}{|l|c|c|c|c|c|c|c|c|c|c|c|c|c|c|} \hline
Query & \#Res & \multicolumn{2}{c|}{Redland }    & Virtuoso&Stardog & HDT&  \multicolumn{4}{c|}{TripleID }     & \multicolumn{4}{c|}{\textbf{Speedup}}     \\ \cline{3-4}\cline{8-15}
               &   & load  & query       &      & &    & load & data & join  & query &   Redland  &     Virtuoso &  Stardog & HDT \\\hline 
       
Q1         &    20,977  & 62.78 & 10.45   &15.76   & 1.62  &   0.65  & 1.86 & 0.29 & -         & 0.41  &  25.49  & 38.43 &3.95&1.58 \\\hline 
Q2          &    1,119,681 & 62.40  & 57.34     & 1173.10 &  821.28  & 3.80   & 1.7 & 0.29 & -         & 0.97  & 59.11  & 1209.38 &846.68&3.92\\\hline 
Q3       &   1,220,456     & 62.52 & 80.9   &1391.87        &880.19 &  4.67& 1.69 & 0.31  & -         & 1.02  & 79.31  & 1364.57 &862.93&4.44 \\\hline 
Q4       &  1,242,627       & 62.72 & 102.09   & 1399.37    & 908.90 &  5.22 & 1.71 & 0.30  & -         & 1.06  & 96.31  & 1320.16 &857.45&4.92\\\hline 
Q5        &    24   & 62.19 & 7.78   & 0.85    &  0.18 &  0  & 1.68 & 0.27 & -         & 0.33  & 23.58 & 2.58 &-& -\\\hline 
Q6        &    18   & 62.35 & 7.75        &0.61  & 0.07 &  0 & 1.69 & 0.29 & -         & 0.33  & 23.48  & 1.85 &-& -\\\hline 
Q7         &   23   & 62.28 & 15.96     &1.86   &  0.13 & 0.58 & 1.68 & 0.3  & -         & 0.38  & 42.00   &4.89 &-&1.52\\\hline 
Q8         & 26,307     & 62.42 & 27.92    & 61.20  & 1.67    &1.15   & 1.68 & 0.3  & -         & 0.45  & 62.04  & 136.00 &3.71&2.55\\\hline 
Q9        &    10   & 62.15 & 50.57      &0.53     & 0.09&1.31 & 1.68 & 0.29 &  0*         & 0.46  & 109.93 & - &-&2.85\\\hline 
Q10        &  0   & 62.61& 51.72        & 3.39  & 0.14 &1.84 & 1.71 & 0.31 &  0*     & 0.49 & 105.56   & 10.93 &-&3.75\\\hline 
Q11        &    98  & 61.59 & 7.73        &0.55   &  0.11& 0.53 & 1.71 & 0.29 & 0*        & 0.35  & 22.09   & 1.58&-&1.5\\\hline 
Q12         &  1,542   & 61.72 & 8.46        & 11.24 & 0.58 & 0.66 & 1.73 & 0.28 & 0*         & 0.39  & 21.69  & 28.82 &1.49&1.69\\\hline 
Q13        &    31,863  & 62.33 & 11.58       &24.71 &  2.99 & 0.73 & 1.72 & 0.3  & 0.03      & 0.41  & 28.24   & 60.27&7.29&1.78\\\hline 
Q14        &  148,213    & 61.55 & 19.18     &73.01   & 20.62  &  1.20 & 1.7  & 0.3  & 0.11      & 0.52  & 36.88   & 140.41&39.65&2.31\\\hline 
Q15         &   5,715    & 63.38 & 9.31        &1.12     & 0.59 &0.61& 1.71 & 0.29 & 0.01      & 0.39  & 23.87   & 2.88&1.51&1.56\\\hline 
Q16         &90,504     & N/A   & N/A         & 2140.42 & 645.56  &  10.67 & 1.72 & 0.31 & 0.32      & 0.76  & -    & 2,816.35 &849.42&14.03\\\hline 
\end{tabular}
\end{table*}

\subsection{Entailment Queries}
We apply our framework to process queries according to entailment rules. Table \ref{tab:entail2} presents  rules used as a query benchmark  \cite{HJ2012} out of 13 $D^*$ rules  \cite{Horst} since the other rules involve only one subquery. These rules are transformed to the queries  which contain two subqueries. Hence, $GPUSearch$ must be called twice.

\begin{table}[htbp]
\centering
\caption{RDFS Entailment rules from \cite{Horst}} 
 
\setlength{\tabcolsep}{0pt}  
\label{tab:entail2} 
 \small
\begin{tabular}{|c|l|l|}\hline
R  & If RDF graph contains & Then ($\Rightarrow$) \\ \hline
(2) & $s$ $p$  $o$   \&\&  & $ s$ \texttt{rdf:type} $D$   \\
&$p$ \texttt{rdfs:domain} $D$ & \\
(3) & $s$ $p$  $o$   \&\&  & $ o$ \texttt{rdf:type} $R$   \\
&$p$ \texttt{rdfs:range} $R$ &\\
(7) & $s$ $p$  $o$   \&\&& $ s$  $q$ $o$   \\ 
& $p$ \texttt{ rdfs:subPropertyOf} $q$  &\\\hline
(5) & $p$ \texttt{rdfs:subPropertyOf} $q$   \&\& & $ p$ \texttt{rdfs:subPropertyOf} $r$   \\
&$q$ \texttt{rdfs:subPropertyOf} $r$  &\\
(9) & $s$ \texttt{rdf:type} $x$    \&\& & $  s$ \texttt{rdf:type} $y$ \\
&$x$ \texttt{rdfs:subClassOf} $y$ &\\
(11) & $x$ \texttt{rdfs:subClassOf} $y$   \&\&  &$  x$ \texttt{rdfs:subClassOf} $z$ \\
& $y$ \texttt{rdfs:subClassOf} $z$&\\\hline
\end{tabular}

\end{table}

\begin{figure}[thbp]
\centering
\subfigure[Step 1: $x$ \texttt{rdfs:subClassOf} $y$]{\includegraphics[width=\columnwidth]{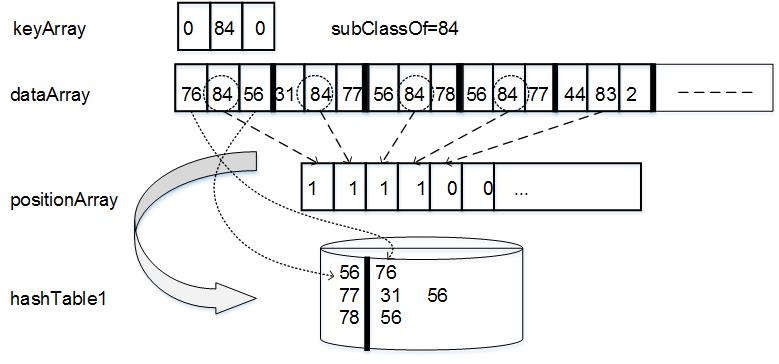}}
\subfigure[Step2: $y$ \texttt{rdfs:subClassOf} $z$]{\includegraphics[width=\columnwidth]{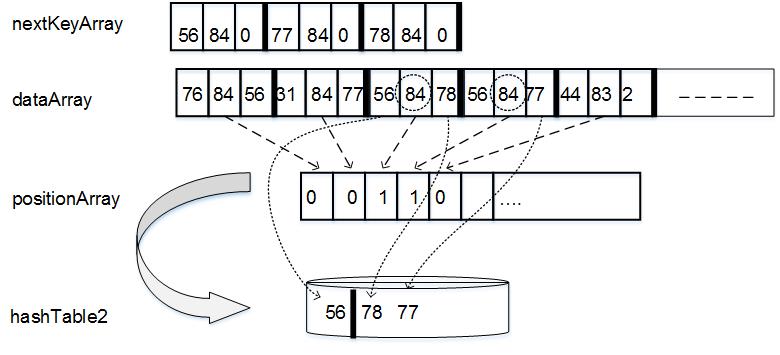}}
\subfigure[Combine results: $x$ \texttt{rdfs:subClassOf} $z$]{\includegraphics[width=3.5in]{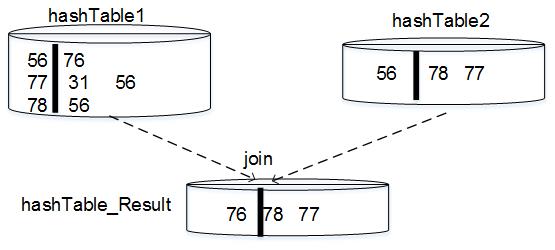}}
\caption{Rule 11's example. }
\label{fig:entailex1}
\end{figure}
 
 Figure \ref{fig:entailex1}  demonstrates an example of computing query for Rule (11)  in Table \ref{tab:entail2}. The rule  implies that
if $x$ is  \texttt{rdfs:subClassOf} $y$  and  $  y$  is \texttt{rdfs:subClassOf} $z$, then 
 $x$ \texttt{rdfs:subClassOf} $z$.
 Suppose that  $dataArray$ contains triples  \{(76, 84, 56), (31, 84, 77), (56,84, 78),
  (56, 84, 77), (44,83,2)\}. 
In Rule (11),  the ``If RDF graph contain'':
 $x$ \texttt{rdfs:subClassOf} $y$  is considered. \texttt{rdfs:subClassOf}  is mapped into Pred ID, e.g., 84. 
 Hence, $keysArray= \{0,84,0\}$, and  the returned $positionArray=$\{1,1,1,1,0\} as shown in  Figure \ref{fig:entailex1}(a).  
 The found triples are gathered into a hash table, called $hashTable1$,
  \{(56,76), (77, (31,56)), (78,56)\}.
Using this hash table, the number of found elements is reduced for the next search.
For the next call of $GPUSearch$, $keysArray=$\{ 56,84,0, 77,84,0, 78,84,0 \} in   Figure \ref{fig:entailex1}(b).
After that $positionArray$  returned is \{0,0,1,1,0\} and
  the matched  subjects and the objects of the found  triples  are put in another hash table, $hashTable2=$ \{(56, (78 ,77))\}.
$hashTable2$ is joined with $hashTable1=$  \{(56,76), (77, (31,56)), (78,56) \}, giving \{(76,78), (76,77)\} in   Figure \ref{fig:entailex1}(c).   

The execution time of  queries in  Appendix
 using  TripleID, HDT, Stardog, Virtuoso, MySQL,    HDT, and TripleID-C  is reported  in Table 
\ref{tab:entail} in column "TripleID", "HDT", "Stardog", "Virtuoso", "MySQL" and "TID/C" respectively.
"TID/C" is TripleID implementation using only the GPU host.  
 In column "\#Res1", the number of the  results   of the first subquery,
for example, in Rule (2),  first query is the ``If  RDF Graph contains" part,  as $p$ \texttt{rdfs:domain} $D$.
 The second query  is to search for all $p$'s that are  previously found in all triples.
For Rule (5), the first query  is  $p$ \texttt{rdfs:subPropertyOf} $D$.
Column "\#Dist1" is the number of distinct results from column "\#Res1".
For column "\#Res2", the number of results is  from the second search. 
Similarly, "\#Dist2" is the number of distinct items from column "\#Res2".
At last, column "All" shows the total combined results from "\#Res" and "\#Res2".

\begin{table*}[phbt]
\setlength{\tabcolsep}{0.5pt}
\def\arraystretch{0.8}
\centering\small
\caption{Execution time  in seconds of  queries according to entailment rules.}
\label{tab:entail}
\addtolength{\tabcolsep}{0.5pt}  
 
 \begin{tabular}[width=\textwidth]{|l||l|c|c|c|c|c||c|c|c|c|c|c|}\hline
Data & Rule & \#Res1 & \#Dist1 & \#Res2 & \#Dist2 &  All     & \multicolumn{1}{c|}{  TripleID} & { HDT }& { Stardog }& { Virtuoso }&{ MySQL}& { TID/C} \\
    \hline
 012347 & R2  & 8,395  & 2,437 & 226,433 & 169 & 169,776   & \textbf{18.09} & 34.15&764.95& 3,073.12 & 4,402.72 	&53.45 \\
       & R3  & 9,589  & 2,505 & 226,099 & 186 & 62,005   & \textbf{2.46}&30.85&740.23& 752.29& 4,904.71 &	35.41\\
       & R5  & 6,545  & 450  & 0      & 0   & 0     & 0.28 & 0.53& \textbf{0.19}&0.23 &  4,177.85 	&6.7   \\
       & R7  & 6,545  & 1,120 & 32,433  & 95  & 22,855    & \textbf{0.55}&38.39&  128.88& 1,776.76 & 6.18 	&20.87\\
       & R9  & 10    & 4    & 1      & 1   & 1      & \textbf{0.19 } &7.03&200.72& 0.25    & 3.36 	&0.03 \\
       & R11 & 26,785 & 4,716 & 87     & 47  & 90      &  1.24  & \textbf{0.65}&2.42&11.99&  53.06 	&69.92 \\\hline
       btc-small
& R2  & 10,185 & 3,596 & 301,680 & 205 & 219,698   & \textbf{23.08} & 49.87&509.12&4,485.87& 7,416.90 	&88.08   \\
              & R3  & 11,438 & 3,592 & 305,591 & 210 & 89,372   & \textbf{3.53 } &45.61 &455.42& 913.18 & 8,125.93 	&77.38  \\
               & R5  & 7,980  & 584   & 0      & 0  & 0        & 0.39  &  0.68&0.76& \textbf{0.22}   & 7,098.16 &	11.58 \\
               & R7  & 7,980  & 1,496 & 57,884  & 100 & 40,622  & \textbf{1.00 }& 53.39&442.34& 1,907.66 & 8.65 	&28.32\\
               & R9  & 10     & 4     & 1       & 1   & 1        &  0.21 & 9.04&657.43& \textbf{0.05}     & 4.55 &	0.26\\
               & R11 & 36,561 & 6,739 & 91      & 49  & 98      &  3.06  &\textbf{0.85}&3.72& 6.22  & 97.22 	&124.58  \\
         \hline 
\end{tabular}
 \end{table*}

In Table  \ref{tab:entail},  after eliminating redundant  results from the first GPU search (Column "\#Res1"),   $keysArray$ size is much smaller.   Only distinct results are sent as inputs to  the second GPU search.
It is obvious that Virtuoso and Stardog can handle large databases very well.
Comparing the  speedup of our GPU version and CPU version, it is obvious that the speedup is up to 42 times.
 Our approach  works well when there  are a lot of intermediate results and
final results, eg. R2,R3,R7 because  of the  simultaneous search from GPU threads. 
If there are very few results for a certain query, then the total execution time  is dominated by memory transfer time as seen in R5, R9, and R11 cases, where Virtuoso or Stardog is  faster. 

\subsection{Effects of Data Transfer Time}

\begin{figure}[thbp]
\centering
\subfigure[Query time and data time]{\includegraphics[width=0.7\columnwidth]{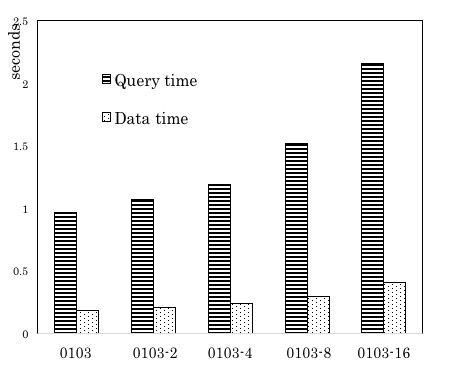}}
\subfigure[Load time]{\includegraphics[width=0.7\columnwidth]{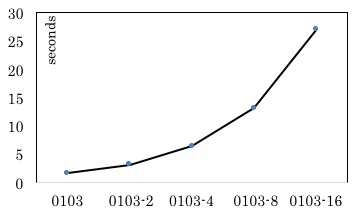}}
\caption{Larger data for query Q2. }
\label{fig:01-03data}
\end{figure}

To observe the scaling aspect, when the number of results to transfer back increases.
Let us consider the data set item `0103' from BTC data set and take Q2 as an example. We double the data `0103', 2 times, 4 times, 8 times, and 16 times, called
0103-2, 0103-4,0103-8 and 0103-16, respectively.
Figure \ref{fig:01-03data}(a) shows the ``Query time''  and ``Data time'' of  Q2 of TripleID.  
The ``Data time'' shows the data transfer of the results back.  For the case 0103-16, the query time is double  from the case for the case 0103, while the data transfer time in this case is about 20\% on average of the  query time. However, the loading time of the data to GPU memory is double as expected in Figure \ref{fig:01-03data}(b).

 \section{Conclusion and Future Work}
\label{sec:conc}
In this paper, we present a framework, \emph{TripleID-Q} based on TripleID format for query processing. 
First,   the conversion from  standard RDF triple format to TripleID   format is performed.
The  subject, predicate, and   object ID files are generated and 
 the TripleID file  which contains   rows of  IDs of subjects, predicates, and objects 
is generated.  
The storage required for all the files is much smaller than the storage used by NT,N3,or RDF file.
The TripleID file is loaded to the  GPU global memory and concurrent search  by GPU threads is done to look for particular subject, predicate, and/or object IDs.
The found triple results are returned.
We demonstrate the application of the search in query processing.
 
The experiments demonstrate various queries where
the intermediate results are filtered, union and/or joined.
While the complexity and the number of results have significant effects in computation time for traditional library, our approach can process the complex query, with large intermediate results in seconds due to the use of large number of simultaneous threads during   searching and joining stages.
Our approach  can give speedup the queries varying from 17-108  times over the traditional RDF query tool. Compared with the above RDF stores, our algorithm can speedup the queries up to hundred times for many union and join operations. When compared with another compact representation, HDT,
the speedup of our algorithm is up to 7 times.
Consider the compactness of the representation. 
The  total ID file size is about 2-4 times  smaller than the original files. 
 It is  only   2 times larger than HDT file size and it is about half size of Stardog RDF store.
On the other hand, TripleID representation is simple so that the conversion time to this format is faster than   HDT's conversion time about 3 times.
The  results show  the trade-off between the compactness, conversion time and query time.

The application of our algorithm to entailment queries also imply the   efficiency.
We gain consistent speedup for these queries over using HDT presentation, 
Stardog, Virtuoso, MySQL.


Our framework relies  on the hash data structure where   three internal hashes for storing subjects, predicates, and objects  are  constructed during   TripleID conversion
The available heap memory limits   the total maximum subjects, predicates, and objects we can store.
This makes the conversion process get killed when it consumes too much memory in the user space. This limitation is eliminated in the next version (demonstrated in  the future version \cite{pisit2018}) where the vector is used in placed of the hash table. Also, if the total triples’ sizes are too large for the available GPU memory, it can also be scaled out to use multiple GPUs to hold several portions of TripleID data similarly  as in \cite{choksuchat2016}.
Streaming process is another solution to overcome this limit. The next implementation will consider streaming operations and external sorting for conversion and   querying.

 .


%


\section*{Acknowledgment}
 This work was supported in part by the following institutes and research programs: The Thailand Research Fund (TRF) through the Royal Golden Jubilee Ph.D. Program under Grant PHD/0005/2554, DAAD (German Academic Exchange Service) Scholarship project id: 57084841, NVIDIA Hardware grant, and the Faculty of Engineering at Kasetsart University Research funding contract no. 57/12/MATE.

\ifCLASSOPTIONcaptionsoff
  \newpage
\fi

\singlespace
\bibliographystyle{IEEEtran}
\bibliography{ref}

 



%
 
%
\begin{IEEEbiography}[{\includegraphics[width=1in,height=1.25in,clip,keepaspectratio]{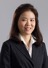}}]{Chantana Chantrapornchai}

 Chantana Chantrapornchai obtained her Bachelor degree  (Computer Science) from  Thammasat University of Thailand in 1991.  She graduated from Northeastern University at Boston, College of Computer Science, in 1993 and University of Notre Dame,  Department of Computer Science and Engineering, in 1999, for her Master and Ph.D degrees respectively.  Currently, she is an associated professor of Dept. of Computer Engineering, Faculty of Engineering, Kasetsart University, Thailand. Her research interests include:  parallel computing,  big data processing, semantic web, computer architecture, and fuzzy logic. She  is also currently with HPCNC laboratory and a principle investigator of GPU Education program at Kasetsart University.
\end{IEEEbiography}

\begin{IEEEbiography}[{\includegraphics[width=1in,height=1.25in,clip,keepaspectratio]{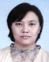}}]{Chidchanok Choksuchat}
 
Chidchanok Choksuchat received the Ph.D. in Computer and Information Science from Silpakorn University. She is currently a lecturer in Information and Communication Technology Programme, Faculty of Science, Prince of Songkla University, Thailand. Her research interests include the CUDA, Java Concurrency, Ontology Engineering, SPARQL Endpoint, Linked Open Data, Data Science Toolkit, R programming and Internet of Things.

\end{IEEEbiography}


%


\vfill


\end{document}